\documentclass[smallextended]{svjour3}       % onecolumn (second format)
\smartqed  % flush right qed marks, e.g. at end of proof
\usepackage{amssymb}
\usepackage{amsmath}
\usepackage[normalem]{ulem}
\usepackage{comment}
\usepackage{xcolor}

\usepackage{geometry}
\usepackage{graphicx}
\usepackage{pdflscape}
\usepackage{hyperref}
\newcommand{\bfY}{{\mbox{\bf{Y}}}}

\begin{document}
% Prepared for Advances in Statistical Analysis
% https://www.springer.com/journal/10182/updates/17805238

\title{\textcolor{black}{Estimation of Player Aging Curves Using Regression and Imputation\thanks{This material is based upon work supported by the U.S.~National Science Foundation under Grant No. CNS-1919554}} 
%\sout{What does not get observed can be used to make age curves stronger: estimating player age curves using regression and imputation} \thanks{This material is based upon work supported by the U.S.~National Science Foundation under Grant No. CNS-1919554}
}
%\subtitle{I don't think\\we have a subtitle}

\titlerunning{Estimating player age curves}        % if too long for running head

\author{Michael Schuckers         \and
        Michael Lopez \and Brian Macdonald} 
        %etc.}


\institute{Michael Schuckers \at
              St. Lawrence University,              Canton, NY \\\email{schuckers@stlawu.edu}   
              \and
    Michael Lopez \at
     National Football League\\
    \email{Michael.Lopez@nfl.com} \and
    Brian Macdonald\\
    Yale University\\
    \email{brian.macdonald@yale.edu}
}

\date{Received: date / Accepted: date}
\maketitle

\begin{abstract}

\textcolor{black}{The impact of age on performance is a fundamental component to models of player valuation and prediction across sport. Age effects are typically measured using age curves, which reflect the expected average performance at each age among all players that are eligible to participate. Most age curve methods, however, ignore the reality that age likewise influences which players receive opportunities to perform. In this paper we begin by highlighting how selection bias is linked to the ages in which we observe players perform. Next, using underlying distributions of how players move in and out of sport organizations, we assess the performance of various methods for age curve estimation under the selection bias of player entry and issues of small samples at younger and older ages. We propose several methods for player age curve estimation, introduce a missing data framework, and compare these new methods to more familiar approaches including both parametric and semi-parametric modeling.}
We then use simulations to compare several approaches for estimating aging curves. Imputation-based methods, as well as models that account for individual player skill, tend to generate lower \textcolor{black}{root mean squared error} (RMSE) and age curve shapes that better match the truth. We implement our approach using data from the National Hockey League.  All of the data and code for this paper are available in a Github repository.

\end{abstract}

%\sout{The impact of age on athlete performance has received attention across sport.
%First, we highlight how selection bias is linked to the ages in which ages we observe players perform. This approach is used to generate underlying distributions of how players move in and out of sport organizations. Second, motivated by methods for missing data, we propose novel estimation methods of age curves by using both observed and unobserved (imputed) data.}

\section{Introduction}
\label{intro}

\textcolor{black}{Accounting for an athlete's age is an issue omnipresent across sport. As examples:
When should a team call up a prospect for a first shot at the big leagues? When to sign a star player to a new contract? How much money will a player be worth at the end of a new deal? When will players tire within a game? And how do the answers to these previous questions differ by player position and athletic traits?}

\textcolor{black}{The effect of age is typically measured using age curves, which reflect the average expected performance across a population of eligible athletes. In a closed population where each athlete played at each age, or where drop-out and eligibility were random, age curves would be straight-forward to estimate using average performance. Unfortunately, data from the sports world is quite the opposite. At both early and late ages, only the most elite athletes perform well enough to have their contributions recorded.}

\begin{figure}[htbp!]
    \centering
    \includegraphics[width=.75\textwidth]{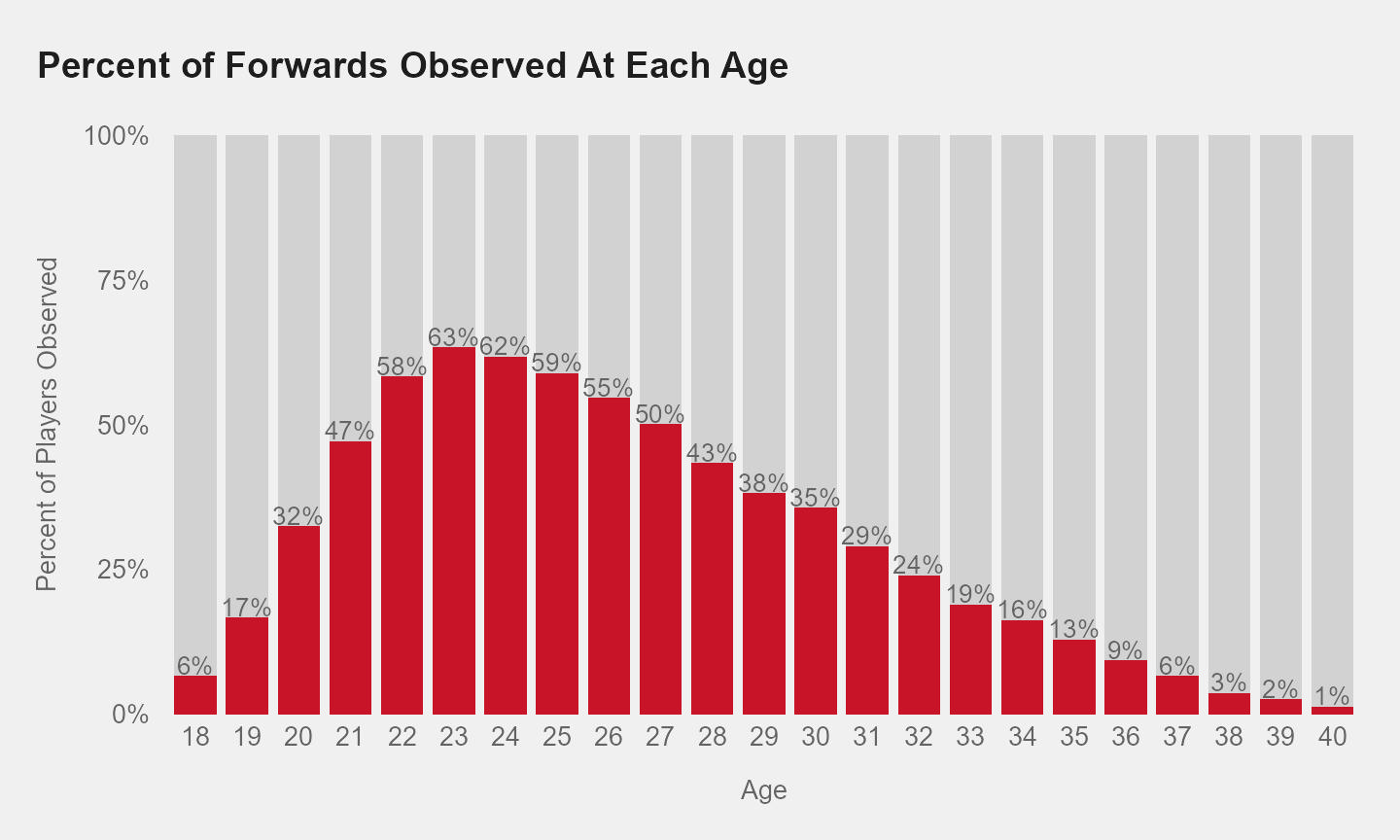}
    \caption{Percent of players that are observed at each age for NHL forwards. The distribution peaks around ages 23-24, and very few players remain in the league through their late 30s.}
    \label{fig:percentobservednhl}
\end{figure}

\textcolor{black}{To demonstrate how age manifests itself in which athletes are observed, Figure \ref{fig:percentobservednhl} shows the percentage of players (n=2276) that are observed by age for forwards in the National Hockey League (NHL). In this instance, observed players are those who  recorded at least 1 game played in a given season, and the population of players reflects any player who played in the NHL from the 1988-89 to the 2020-21 seasons.}

\textcolor{black}{In Figure \ref{fig:percentobservednhl}, the rate of observed data for NHL forwards peaks around ages 23-24, the time in which about 63\% of players are observed. At age 18, and by the time players reach age 36, fewer than 1 in 10 (approximately 9\%) of players are observed. }

\begin{figure}[htbp]
    \centering
    \includegraphics[width=.45\textwidth]{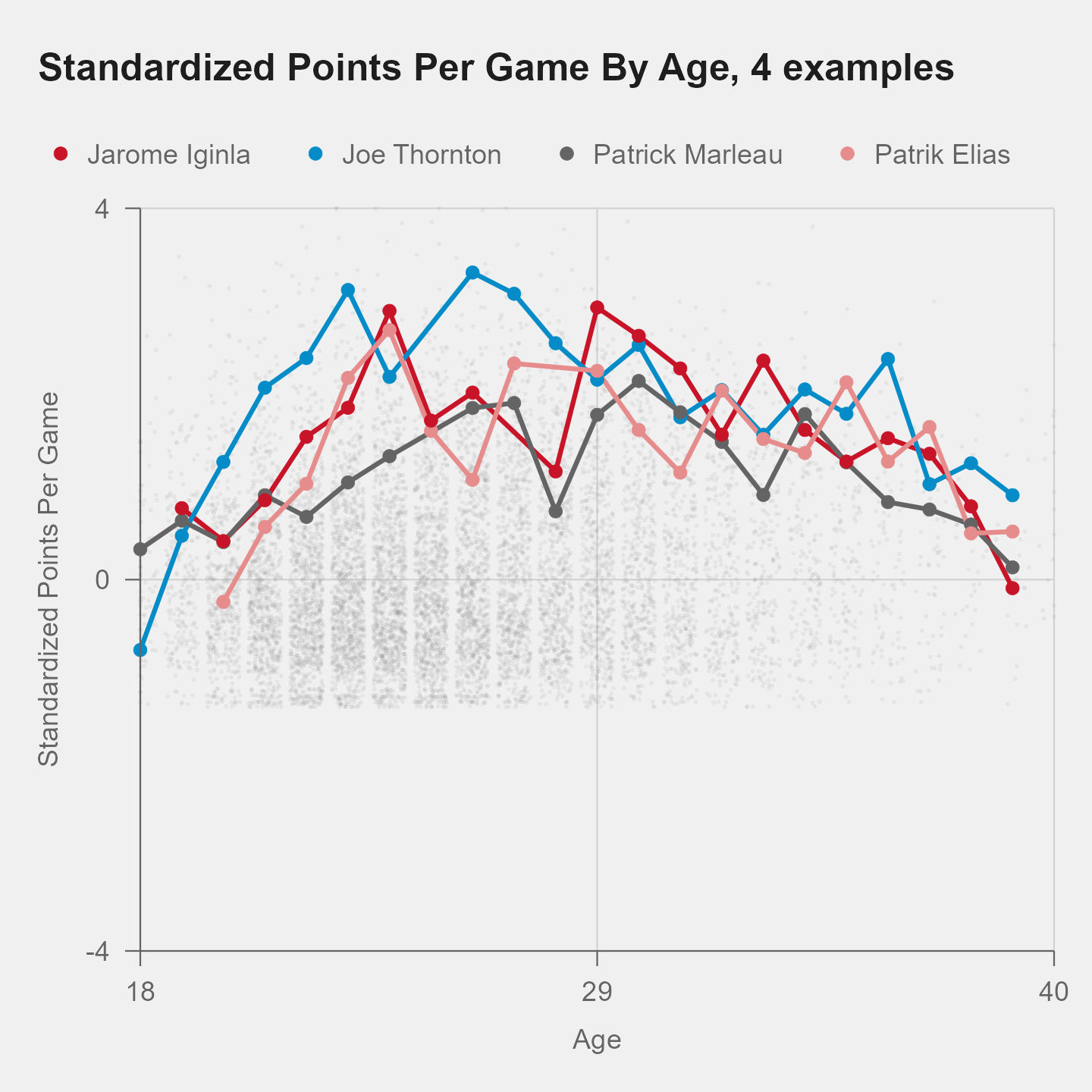}
    \includegraphics[width=.45\textwidth]{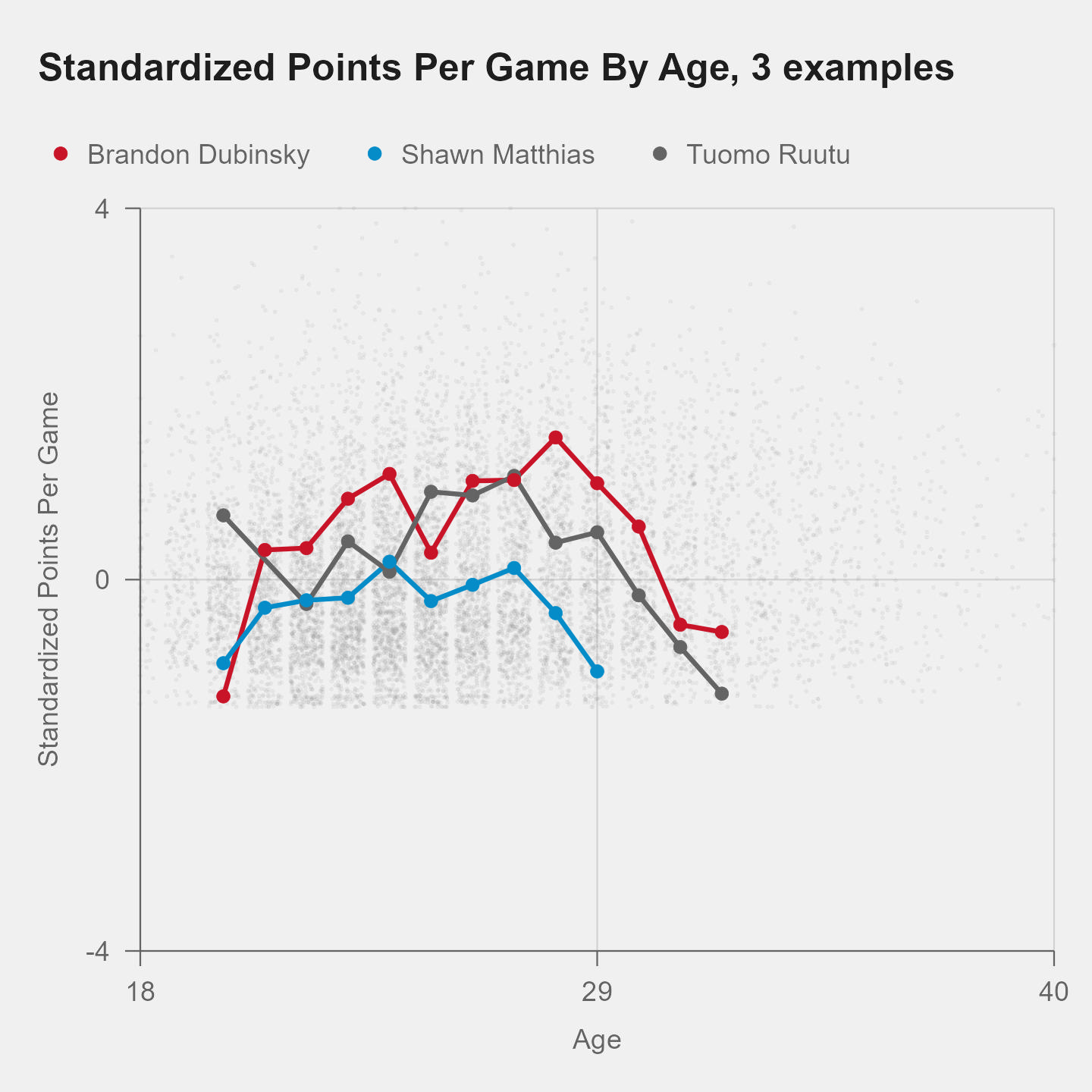}
    \caption{Standardized points per game for four examples of good forwards who played in the NHL until at least there late 30s (left), and three examples of good NHL forwards did not stay in the NHL through their late 30s. Individual player statistics over time tend to follow an inverted U-shaped trend, increasing from the mid to late 20s, and decreasing thereafter. For the players on the right, once their production fell, they exited the league, and have no observed data in their late 30s that would factor into the curve generated in Figure \ref{fig:bootstrap}.}
    \label{fig:exampleplayers}
\end{figure}

\textcolor{black}{Given both physical and mental maturity ranges, age curves are often observed to have a concave downward shape, where peak performance lies somewhere in the middle of a career, with drop-offs at both the beginning and end stages of a career. Anecdotally, this tends to match the performances of several athletes.} \textcolor{black}{In Figure \ref{fig:exampleplayers}, for example, we highlight observed data for four example players who were in the league at least until their late 30s (left) and three players who were in the NHL through their late 20s and early 30s (right). These data for individual players initially exhibit the same general trend -- increasing through their mid-to-late 20s, and decreasing in their 30s.}

% \begin{figure}
%     \centering
%     \includegraphics[width=.45\textwidth]{img/StandardizedPointsPerGameByAgeNHLForwards.jpg}
%     \caption{Standardized points per game by age for NHL forwards between the 1995-96 and 2018-19 seasons, along with a cubic spline model fit to the data. The spline model does not decrease as it should for older ages because of selection bias: the only players that are observed at older ages are very good players. }
%     \label{fig:observedwithsplinewithoutimputation}
% \end{figure}

\textcolor{black}{Given the selection bias of player entry and exit from leagues in Figure \ref{fig:percentobservednhl}, methods that insufficiently account for player skill may not fully account for the impact of age.} 

\begin{figure}[htbp!]
    \centering
    \includegraphics[width=.45\textwidth]{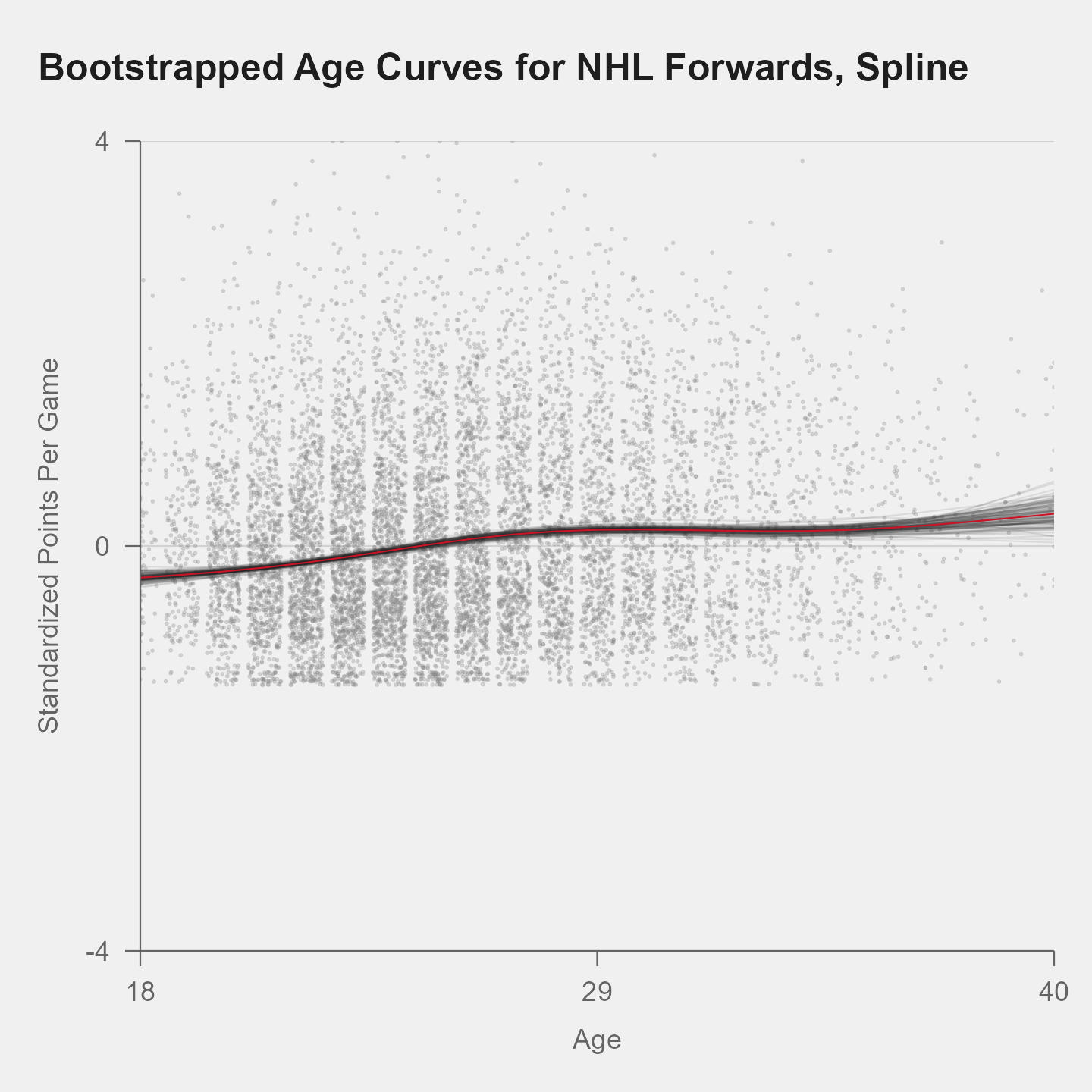}
    \includegraphics[width=.45\textwidth]{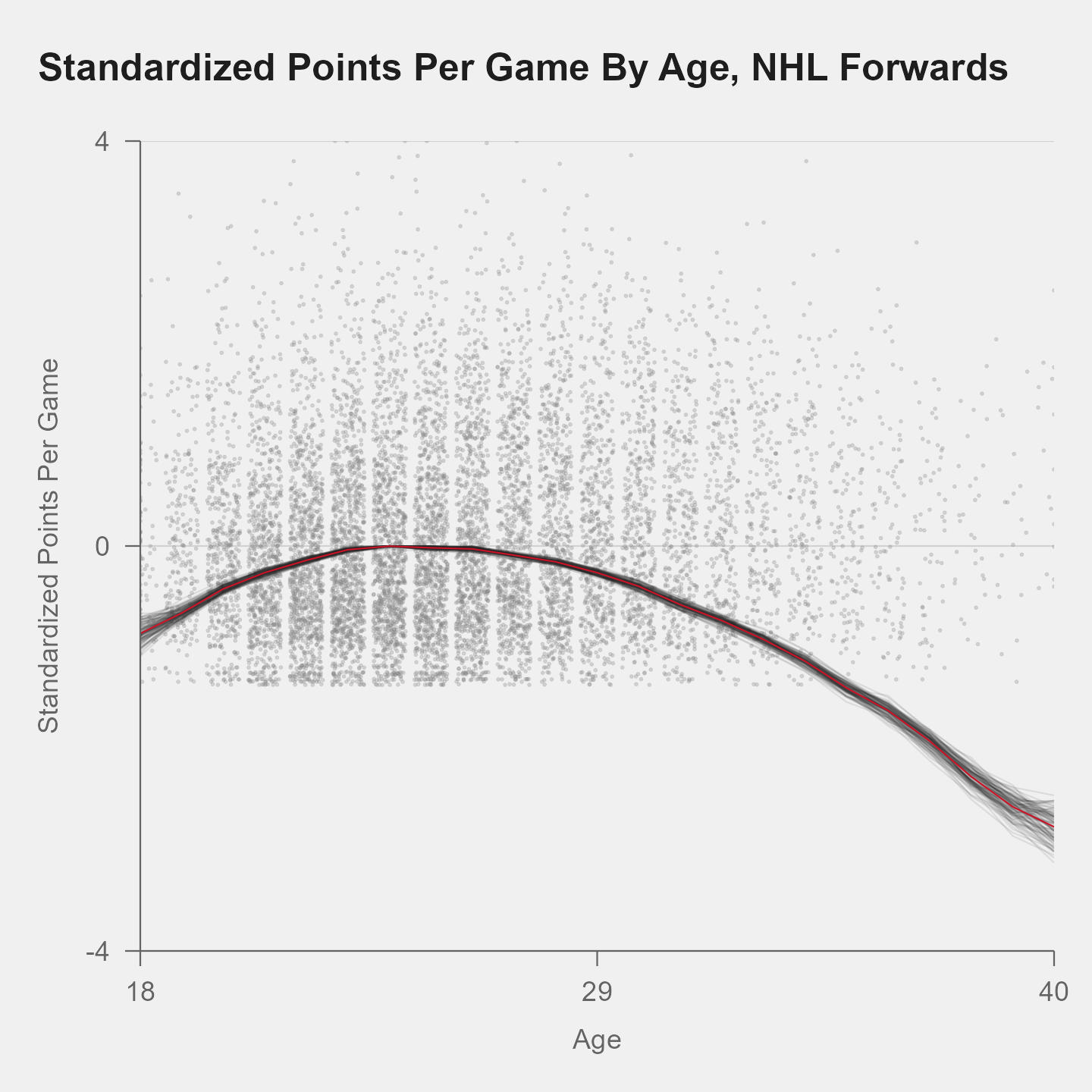}
    \caption{Standardized points per game by age for NHL forwards between the 1995-96 and 2018-19 seasons, along with 100 bootstrapped age curves using the natural spline regression model without player effects (left) and Delta Method (right) with the NHL Forwards data. The naive spline model does not decrease as it should for older ages because
of selection bias: the only players that are observed at older ages are very good players.  For both methods, variation in the age curves increases for older ages where there are fewer observed data.}
    \label{fig:bootstrap}
\end{figure}

As an example of what can go wrong, the left side of Figure \ref{fig:bootstrap} shows standardized points per game for National Hockey League (NHL) forwards, along with 100 age curves derived using a cubic spline using only observed data and no player effects for estimating the age curves for 100 bootstrapped samples of the NHL forwards data. There seems to be a slight bump at 28 or 29, but then the curve starts \textit{increasing} \textcolor{black}{around age 33}, which is the opposite of what we would expect from an age curve. Because average and below average players tend drop out, the only players competing in their late 30s tend to be very good players.  On the right side of Figure \ref{fig:bootstrap}, age curves generated using the Delta Method \cite{mlicht2009} do show more reasonable looking shapes, tending to peak around age 26.  Both sets of curves in Figure \ref{fig:bootstrap} show low variance for some ages (e.g. mid 20s), but variance increases for younger and older ages as fewer observed data are available.

\textcolor{black}{The goal of this paper is to identify the statistical tools best equipped to estimate the shape of age curves. Below we generalize the Delta Method to allow for non-zero maxima of a mean age curve and refer to this generalization as the Delta Plus method. Since the form of mean age curves is assumed to be concave downward but the specific form of these curves can vary, we consider a range of different estimation approaches for mean age curves including models with fixed effects and models with random effects as well as different data to be included in the estimation process.  Specifically, we consider regression spline models, a novel smoothed quantile approach, and a quadratic model to compare against the Delta Plus method. For players whose performance is not observed since they did not play in the top league, e.g. the NHL in ice hockey or the NBA in basketball, we consider imputation of these values to improve model performance. We consider approaches that include imputation that is either truncated to take values below some threshold or is not truncated.  As we will see in Section \ref{sec:methodologies}, we combine a variety of these methods.}

\textcolor{black}{To assess how well these models perform we consider two methods for evaluating model performance given a known underlying mean age curve in Section \ref{sec:evaluation}.  The first is the root mean squared error (RMSE) at each player age.  This method directly assesses how well each estimation method performs by looking at the error distance at each age.  The second assessment method is to use shape based distance (SBD) formulated by \cite{paparrizos2015k}.  This approach quantifies the similarity of curve shapes by taking a normalized cross-correlation approach.  Results for both assessments methods are given for simulated data, and then methods are compared using NHL data. Overall, we find methods that account for individual player skill tend to outperform ones that do not, as assessed by lowest RMSE and highest SBD.}  Data and code used in this paper can be found at \href{https://github.com/schuckers/playeraging}{https://github.com/schuckers/playeraging}.

\section{Methodologies} \label{sec:methodologies}

For consistent terminology, we begin with the following notation. Let $Y_{it}$ to represent the performance value of player $i$ at age $t$. We assume discrete observations at each age (one per year), though it is possible to treat age in a more granular way.  

A basic model is: 
\begin{equation}
    Y_{it} = g(t) + f(i, t) + \epsilon_{it}
    \label{eq:model}
\end{equation}
where $g(t)$ is the average performance at age $t$ for all players, $f(i, t)$  represents a possible performance adjustment at age $t$ for player $i$, and $\epsilon_{it}$ is the model error at age $t$ for player $i$.  \textcolor{black}{The function $g(t)$ is the primary focus of the estimation and evaluation methods in this paper.  In the models below we will consider both fixed and random effects version of $f(i,t)$ to estimate player effects.}

%In general we will not assume a parametric form for $g(t)$ and $f(i, t)$, but assume that they are smooth functions over age $t$.  The ultimate goal of this paper is to develop and compare methods for the estimation of $g(t)$.  

\subsection{Related literature}

Broadly, current approaches can be split based on assumptions for \textcolor{black}{1)} estimating $g(t)$ (parametric, semi-parametric, or non-parametric) and \textcolor{black}{2)} whether or not to model age specific curves, $f(i, t)$. 
Table \ref{tab:table1} summarizes how several authors have modeled age effects\footnote{\textcolor{black}{MLB stands for Major League Baseball and NBA stands for National Basketball League}}. Columns for author, sport or league, sample of players, and method. Articles in Table \ref{tab:table1} are arranged by a rough categorization of approach for modeling the age term and then alphabetically by author. 

\begin{table}[htbp!]
  \begin{center}
    \caption{Summary of work on age curves}
    \label{tab:table1}
    \begin{tabular}{l|c|c|c|c|c} 
      \textbf{Paper} & \textbf{Sport/League} & \textbf{Unobserved?} & \textbf{Model} \\
      \hline \hline
      Schulz et al\cite{schulz1994relationship} & MLB & No & Average   \\
      Lichtman\cite{mlicht2009} & MLB & No & Fixed effects (Delta Method)   \\
      Tulsky\cite{Tulsky2014} & NHL & No & Fixed effects (Delta Method)  \\
      Albert\cite{albert2002smoothing}  & MLB & No&  Fixed effects (Quadratic)   \\
      Bradbury\cite{bradbury2009peak} & MLB & No &  Fixed effects (Quadratic)  \\
      Fair\cite{fair2008estimated} & MLB, others & No &  Fixed effects (Quadratic) \\
      Villareal et al\cite{villaroel2011elite}&Triathlon &No&Fixed effects \\
      Brander et al\cite{brander2014estimating} & NHL & Yes & Fixed effects (Quadratic, Cubic)  \\
      Tutoro\cite{tutoro2019} & NHL & No & Semiparametric \\
      Judge\cite{jjudge2020b} & MLB & Yes & Semiparametric \\
      Wakim and Jin\cite{wakim2014functional}  & MLB, NBA & No & Semiparametric\\
      Vaci et al\cite{vaci2019large} & NBA &  No  & Fixed, random effects \\
      Lailvaux et al\cite{lailvaux2014trait} & NBA & No &  Random effects  \\
      Berry et al\cite{berry1999bridging} & MLB, NHL, Golf & No & Random effects   \\
      Kovalchik and Stefani\cite{kovalchik2013longitudinal} & Olympic & No & Random effects    \\
     \hline 
    \end{tabular}
  \end{center}
\end{table}

A most naive approach assumes $f(i, t)= 0$ and $\hat{g}(t) = \sum_{i=1}^{n} Y_{it}/ n$ for all $t$, as in \cite{schulz1994relationship}, such that the average of observed $Y_{it}$ is sufficient for estimating $g(t)$. Such an approach would only be valid if players were chosen to participate in sport completely at random, making it too unrealistic for professional sport. As in the left of Figure \ref{fig:bootstrap}, we can see that this approach yields results that are not credible.

A common parametric assumption \cite{albert2002smoothing}, \cite{brander2014estimating}, \cite{bradbury2009peak}, \cite{fair2008estimated}, \cite{villaroel2011elite} is that $Y_{it}$ is quadratic in age. If $g(t)$ is quadratic, performance ``peaks'' at some age, with players improving until this peak and eventually declining after the peak. Quadratic age curves are also symmetric across the peak performance age. \cite{brander2014estimating} and \cite{villaroel2011elite} also consider a cubic age effect, while \cite{albert2002smoothing} uses a Bayesian model and weighted least squares, with weights proportional to player opportunity. \cite{fair2008estimated} do not assume that quadratic form of $f(i, t)$ is symmetric. The Delta Method \cite{mlicht2009} is a modified fixed effects model using different subsamples at each consecutive pairs of ages, \textcolor{black}{and} is expanded upon in Section \ref{sect::Delta}. 

A second suite of approaches assumes a semiparametric approach for estimating age curves, including spline regression techniques \cite{wakim2014functional} and the broader family of generalized additive models \cite{jjudge2020b}, \cite{tutoro2019}. Either approach is more flexible in their ability to pick up on non-linear patterns in age effects. A final assumption models age effects via individual age curves $f(i, t)$, as in \cite{berry1999bridging} and \cite{lailvaux2014trait}. 

In all but two examples above, authors use observed data only to make inferences on the impact of age. Conditioning on players being observed at each age potentially undersells the impact of age; only players deemed good enough to play will record observations. The two exceptions to this assumption are \cite{brander2014estimating}, who extrapolate model fits to players not observed in the data, and \cite{jjudge2020b}, who uses a truncated normal distribution to estimate errors.  Importantly, \cite{jjudge2020b} confirms that if dropout rate is linked to performance, that effectively shifts the age effect downward, relative to surviving players.

\subsection{Notation}

Our focus in this paper will be on the estimation of $g(t)$, the average aging curve for all players.  Because not all of the players will be observed in a given year of their career (due to injury, lack of talent, etc), we create $\psi_{it}$, an indicator for if $Y_{it}$ was observed for player $i$ at a given age $t$, where $t = t_1, \ldots,t_K$.  That is, 
\begin{eqnarray}
\psi_{it}=\left\{
\begin{array}{ll}
1 & \mbox{if $Y_{it}$ is observed, and}\\
0 & \mbox{otherwise}.\\
\end{array}
\right.
\end{eqnarray}
we let $t_1$ be the youngest age considered and $t_K$ will be the oldest age considered. \textcolor{black}{In this notation, $K$ is the total number of years of performance data for a set of players.  For example, if we consider player performance from age $20$ to $32$, then $t_1=20$, $t_K=32$, and $K=13$.}

%\textcolor{black}{
%\sout{Similarly we will have players who we would like to include in our analysis but their careers are not yet complete.  To that end we let $\phi_{it}$ is an indicator of if $Y(t)_{it}$ is observable.  For our application to NHL data $\phi_{it}$ is an indicator that the performance could have happened by October 2020, when the data for this project was collected.}}    

\begin{comment}
\begin{eqnarray}
\phi_{it}=\left\{
\begin{array}{ll}
1 & \mbox{if $Y_{it}$ is observable, and }\\
0 & \mbox{otherwise}.\\
\end{array}
\right.
\end{eqnarray} 
\end{comment}

\textcolor{black}{For the methods below, we also add the following definitions and notation.  The data for a given player will be represented by the vector of values $\bfY_i=(Y_{it_1}, 
\ldots,Y_{it_K})^T$ and the observed subset of those values will be denoted by the vector of values $\bfY_i^{obs}=(Y_{it} \mid \phi_{it} =1)^T$.
Let $\bfY^{obs} =((Y_{1}^{obs})^T, (Y_{2}^{obs})^T, \ldots, \linebreak[3]
(Y_{N}^{obs})^T)^T$ be the vector of all observed values, while $\bfY = (Y_{1}^T, Y_{2}^T, \ldots, Y_{N}^T)^T$.}

\subsection{Estimation Methods}

In this section we describe several current and novel approaches to estimation of the mean aging curve, $g(t)$.  We begin by describing the de facto standard methodology in the sport analytics literature, the Delta method, \cite{mlicht2009}, and an extension that we call Delta Plus.  Below we 
outline some facets of our proposed approaches to the problem of estimation of $g(t)$.  Roughly these methods breakdown into  the general approach to estimation of $g(t)$, the data to be used for model fitting and the additional fixed or random effects terms.  To help the reader we develop a notational shorthand for combining these facets that is \textit{method:data:effects}.

\subsection{Mean aging curve estimation}

For this paper we consider four approaches to estimation of the mean player aging curve.  The first is the non-parametric Delta method and a simple extension of this approach which we call \textit{delta-plus}.  The Delta Method which been discussed by \cite{jjudge2020a}, \cite{tutoro2019} and \cite{mlicht2009}, is commonly used in practice.  The second approach that we consider is a natural spline regression approach which we denote by \textit{spline} and the third is a quadratic model \textit{quad}.  Finally, we propose a novel \textit{quantile} methodology that utilizes information about the ratio of observed players to observable players at a given age.    

\begin{table}[b]
    \centering
    \caption{Summary of Estimation Methods}  \label{tab:estim}
    \begin{tabular}{l|l|c|c}
    \hline  Method Name & Model formulation &Estimation of $g(t)$ & Imputation \\
         \hline
 delta-plus & Non-parametric  & Piecewise & No\\
  spline:obs:none&
    $s(t)$ & Natural Splines & No\\
   spline:obs:fixed& $s(t) + \gamma_{0i}$& Natural Splines & No\\
  spline:trunc:fixed& $s(t)+\gamma_{0i}$& Natural Splines & Truncated\\
    spline:notrunc:fixed &$s(t)+\gamma_{0i}$& Natural Splines & Not Truncated\\
  quant:trunc:fixed& $\zeta(t)+\gamma_{0i}$ & Quantile Approx. & Truncated \\
  quant:obs:none & $\zeta(t)$ &Quantile Approx. & No\\
  quad:trunc:fixed& $\gamma_0+\gamma_{0i} +\gamma_1 t + \gamma_2 t^2$& Quad. Linear Model& Truncated\\
  spline:trunc:random-quad&$s(t)+g_{0i} + g_{1i} t + g_{2i} t^2$ & Natural Splines & Truncated\\
  spline:trunc:random-spline& $s(t) +g_{0i} + \xi_i(t)$ & Natural Splines & Truncated\\
    \hline
    \end{tabular}
\end{table}

\subsubsection{Delta Method} \label{sect::Delta}

The Delta Method is an approach to estimation that focuses on the maxima of $g(t)$ over $t$. The basic idea is to estimate the year of year change in the average player response curve by averaging among only the players that we observed in both years.  A benefit of this approach is that it implicitly adjusts for a player effect by only using those players who have appeared in both years $t_k$ and $t_k + 1$.  Further, by simply calculating year over year averages, this approach is hyper-localized.  However, the requirement that $Y_{t_k}$ and $Y_{t_k +1}$ are both observed (i.e. that $\psi_{i t_k}=\psi_{i t_k+1}=1$) can be limiting in sports where there is significant drop-in/drop-out across years or seasons.

To develop the delta method, let 
\begin{equation}
    \delta_{t_k} = \overline{Y}_{i*_k,t_{k}+1} - \overline{Y}_{i*_k,t_{k}}
\end{equation}    
    where $$\overline{Y}_{i*_k,t_{k}} = \frac{1}{n*_{t_k}} \sum_{i} Y_{i \in i*_k, t_{k}},$$ 
$i*_k=\left\{ i \mid \psi_{it_k} = \psi_{it_{k}+1}=1\right\}$, and $n*_{t_k} = \vert i*_k \vert$ is the number of elements in $i*_k$.  Thus,  $i*_k$ is the set of players whose performance value was observed at age $t_k$ and $t_k +1$.  Traditionally, the delta method is standardardized 
\begin{equation}
    \hat{g}(t) = \delta_{t} - \max_k \delta_{t_k}
\end{equation}
 for $t = t_1, \ldots, t_K$ so that the largest value of $\hat{g}(t)$ is zero. 
To date the Delta Method has proved effective at estimation of the maxima of $g(t)$.  
See \cite{tutoro2019} and \cite{jjudge2020a} for details on some evaluation of the performance of the Delta method.  This performance relative to other methods that use only observed data is likely due to this focus on year over year differences, which are susceptible to small sample size issues for older ages that can cause non-smooth age curves.  Older ages are not (as big of) an issue for regression techniques (e.g. spline regression), which use information from nearby ages and result in a smooth age curve even when sample sizes are small.

\subsubsection{Delta Plus Method}
One drawback of the Delta Method as described above is that the maximal value for $\hat{g}(t)$ is forced to take the value zero.  In part, this has been the case because the focus of estimation was on the age of maximal performance not on the estimation of the full aging curve; though the application of the Delta Method has taken on the latter function.  \textcolor{black}{The Delta Plus method is thus a more flexible version of the Delta Method allowing for maxima that are different from zero.}Using the notation from above we define the estimation of $\mu_t$ as the following: 
\begin{equation}
    \hat{g}(t) = \delta_{t} - \max_k \delta_{t_k} + \max_{k} \overline{Y}^{obs}_{\cdot t_k},
\end{equation}
where 
$\overline{Y}^{obs}_{\cdot t_k} = \frac{1}{n_{t_k}} \sum_{i : \psi_{it_k}=1} Y^{obs}_{it_k}$ is the average of the $n_{t_k}$ observed values at age $t_k$.  Below we will refer to this method as \textit{delta-plus}.  We only use the Delta Plus Method in our evaluation below since it offers a more flexible approach than the Delta Method.

\textcolor{black}{\subsubsection{Quadratic approach}
As the name implies, this methodology assumes that the mean aging curve is quadratic in terms of a player's age.  In our shorthand we will use `quad' and the model will be written as $g(t) = \gamma_0 + \gamma_1 t + \gamma_2 t^2$.  Some authors including \cite{fair2008estimated}, \cite{albert2002smoothing} and \cite{brander2014estimating} have assumed that $g(t)$ has this particular functional form.  So we include this approach for comparison.}

\subsubsection{Spline approach}
For this spline approach (spline), we utilize flexible natural basis spline regression and apply them to the $Y_{it}$'s, the player performance data.  \textcolor{black}{Spline regression methods do not assume a parametric form for the relationship between the response and the predictor.  Consequently, they allow for pliability of the model fit.}  Our spline regression use age, $t$, as a predictor and the performance metric $Y_{ik}$ as the response.  In particular, we use the \texttt{s()} option with 6 degrees of freedom in the \texttt{mgcv} package from \texttt{R}.  In our shorthand for these methods we call this approach \textit{spline} and use $s(t)$ to denote a spline model for the mean aging curve.

%\textcolor{black}{
%\sout{As the name implies, this methodology assumes that the mean aging curve is quadratic in terms of a player's age.  In our shorthand we will use `quad' and the model will be written as $g(t) = \gamma_0 + \gamma_1 t + \gamma_2 t^2$.  Some authors including \cite{fair2008estimated}, \cite{albert2002smoothing} and \cite{brander2014estimating} have assumed that $g(t)$ has this particular functional form.  So we include this approach for comparison.}}

\subsubsection{Quantile approach}
In this method we aim to estimate $g(t)$ by utilizing the distribution and number of the observed values at each age $t$.  We observe at each age $t$ some fraction of the players which is reasonably assumed to be a truncated sample from a larger population. Let $n_t$ be the number of players whose performance is observed at age $t$ from the population of $N_t$ players whose performance was observable at any age.  If we know the fraction of players relative to the larger population (and the form of the distribution) then we can map percentiles in our sample to percentiles in the larger population.  

For example, suppose we had a corpus of $N_{32} =1000$  players and that at age 32 we observed $n_{32}=400$ of them.  The 75th percentile of observed metrics among the age 32 players might be reasonably used as an estimate for the 90th percentile of the population of 1000.  Explicitly this is because the $100^{th}$ best $Y_{it}$ among the combined observed can be assumed to be the $100^{th}$ best $Y_{it}$ among the $n_{32} =400$ observed and the $N=1000$ unobserved $Y_{it} 's$.   Similarly the 25th percentile of the above example would be the 70th percentile in the population. Below we assume a Normal distribution and use the additivity of the Normal distribution to obtain estimates for $g(t)$ at each age $t$.  
More generally, we can calculate
the $q \cdot 100^{th}$ percentile from among the $n_t$ observed values.  Call that value $\nu_t$.  The value of $\nu_t$ is approximately the $G_t = \left(1-\frac{n_t}{N_t}(1-q)\right) \cdot 100^{th}$ percentile from the population of values at age $t$.
From our example above the $0.90=1-\frac{400}{1000} (1-0.75).$

From $G_t$ we can estimate the mean of the full population, $g(t)$ at age $t$ via: $ \hat{\zeta}_t = \nu_t - \Phi^{-1}(G_t) \hat{\sigma}_t$ where $\Phi(~)$ is the cumulative density of a standard Normal distribution.  From $\hat{\zeta}_t$ we can estimate other percentiles as long as we can assume Normality and we have a reasonable estimate of the standard deviation, $\hat{\sigma}_t$.  Our justification for assuming Normality notes that most performance metrics are averages or other linear combinations of in-game measurements and that the Central Limit Theorem applies to these linear combinations.

For estimation of $\sigma_t$, the standard deviation of the population at age $t$, we can estimate the standard deviation at age $t$, denote that by $s_{t}$, and adjust based upon the proportion of truncation.  Note that the variance of a truncated Normal is always less than the variance of an untruncated one with similar underlying variance.  So then we use $\hat{\sigma}_t =  \frac{s_t}{\theta_t}$ where $\theta_t$ is the ratio of truncation from a standard Normal for these data at age $t$. To obtain $\theta_t$ we use the \texttt{vtruncnorm} function \textcolor{black}{to calculate the variance of the truncated Normal distribution relative to that of a similar untruncated Normal distribution} via the \texttt{truncnorm} library in \textit{R} \cite{rsoftware}.   For our methodology shorthand we will use \textit{quant} for this approach and denote the quantile based aging curve approach as $\zeta(t)$.

\subsection{Data Imputation Methods}
Another modeling choice to be made when working with missing data is whether or not to impute the missing values.
We consider three possible options for imputation of responses, $Y_{it}$'s.  The first option is simply to use \textcolor{black}{only} the observed data %\textcolor{black}{\sout{and only the observed data}}. 
When we do this, we will use 
%\textcolor{black}{\sout{use} denote this by}
\textit{obs} as part of our shorthand.  The second and third options involve imputation of the unobserved but observable data.  
%\textcolor{black}{\sout{, that is the set of values $\{ Y_{it} \mid \psi_{it}=0, \phi_{it}=1 \}.$}}
In the second option\textcolor{black}{, denoted by }\textit{trunc}, we impute values for $Y_{it}$ with truncation\footnote{\textcolor{black}{Imputation with truncation limits the range of values that generated realizations can take.  For example, if $X \sim N(0,1)$ is truncated below at $-2$ then the realized values of $X$ could only take values in $(-2,\infty)$ whereas untruncated realizations could take values in $(-\infty, \infty)$. }}.  In the third option, \textcolor{black}{denoted by} \textit{notrunc} we impute values without truncation.  Below we describe our algorithm for imputation.

In this general imputation algorithm, we first estimate a naive aging curve using a regression approach on only the observed data.  Then we use that estimated curve to generate imputed values for the unobserved values, those \textcolor{black}{$Y_{it}$} with $\psi_{it}=0$,  ensuring that these \textcolor{black}{imputed} 
%\textcolor{black}{\sout{unobserved}} 
values are below a 
%\textcolor{black}{\sout{smoothed}} 
boundary 
%\textcolor{black}{\sout{upper}} 
threshold at each age which is defined by a %\textcolor{black}{\sout{lower}} 
percentile of the observed values.  Our default percentile is the $75^{th}$ percentile.  \textcolor{black}{Thus, with this approach imputed values for a given $Y_{it}$ are forced to fall below a }The choice of the $75^{th}$ percentile is arbitrary, though we considered several choices, in particular, the $20^{th}$ and $50^{th}$ percentiles and found that using the $75^{th}$ percentile produced better estimation performance.  It is sensible that there is an upper bound on the performance of a player who is not in a top league would have; otherwise they would be in the top league.  We then refit our estimated model using both the observed and imputed values and use this new estimated curve to generate a second set of imputed values.  It is this second set of imputed values that we report as our estimate of the the mean aging curve $g(t)$.   The impetus behind the second imputation and curve estimation is to wash out any initial impact on estimated curve which is based solely on the observed values.  

For the algorithm below:
\begin{enumerate}
\item Fit the model $Y^{obs}_{it} =\eta^{'}_{it}+\epsilon_{ij}$ for just the fully observed data, i.e. $Y_{it}$  where $\psi_{it}=1$.  Calculate the estimated standard deviation of the $\hat{\epsilon}_{ij}$, call that $\sigma_0$.   

\item  Estimate a smoothed boundary via splines of performance for players in the NHL based upon the $q^{th}$ percentile of players at each age, say the $75^{th}$ percentile.  Call the boundary value $\tilde{\beta}_{t}$.  Note that for imputation without truncation, \textit{notrunc}, $\tilde{\beta}_{t} = \infty$.

\item Simulate values for the $Y_{it}$ with $\psi_{it}=0$ from a Normal distribution with mean $\eta^{'}_{it}$ and standard deviation $\sigma_0$ but truncated so they are not larger than $\tilde{\beta}_{t}$.  That is the range of possible values for these $Y_{it}$ is $(-\infty, \tilde{\beta}_{t})$, we will denote this by $Y_{it}^{imp} \sim TN (\hat{\eta}^{'}_{it},  \hat{\sigma}_{0}^2, -\infty, \tilde{\beta}_{t})$ where $Y \sim TN(\mu, \sigma^2, a, b)$ denotes a random variable $Y$ with a Truncated Normal distribution with mean $\mu$, variance $\sigma^2$ and takes values such that $a \leq Y \leq b$.  Call these simulated values $Y^{imp}_{it_k}$.

\item Fit $Y^{full}_{it} = \eta_{it}+\epsilon_{it}$ using  \linebreak[3] $$\bfY^{full} = \left\{ Y^{obs}_{it_k}, Y^{imp}_{it_k} \mid  i=1,\dots, N,  k=1,\ldots, K\right\}.$$ 

\item We next impute the original missing values again.  We do this because the first set of imputed values, the $Y^{imp}_{it}$'s,  was based upon predicted values from a model that used only fully observed data.  So then we imputed again, this time using $Y_{it}^{imp} \sim TN(\hat{\eta}_{it}, \hat{\sigma}^2_0, -\infty, \hat{\beta}_{t})$.

\item We then refit the model above using 
$Y^{full}_{it} = g(t) + f(i,t) + \epsilon_{it}$ based upon the imputed data from the previous step.  
\end{enumerate}

\begin{figure}[htbp]
    \centering
    \includegraphics[width=.45\textwidth]{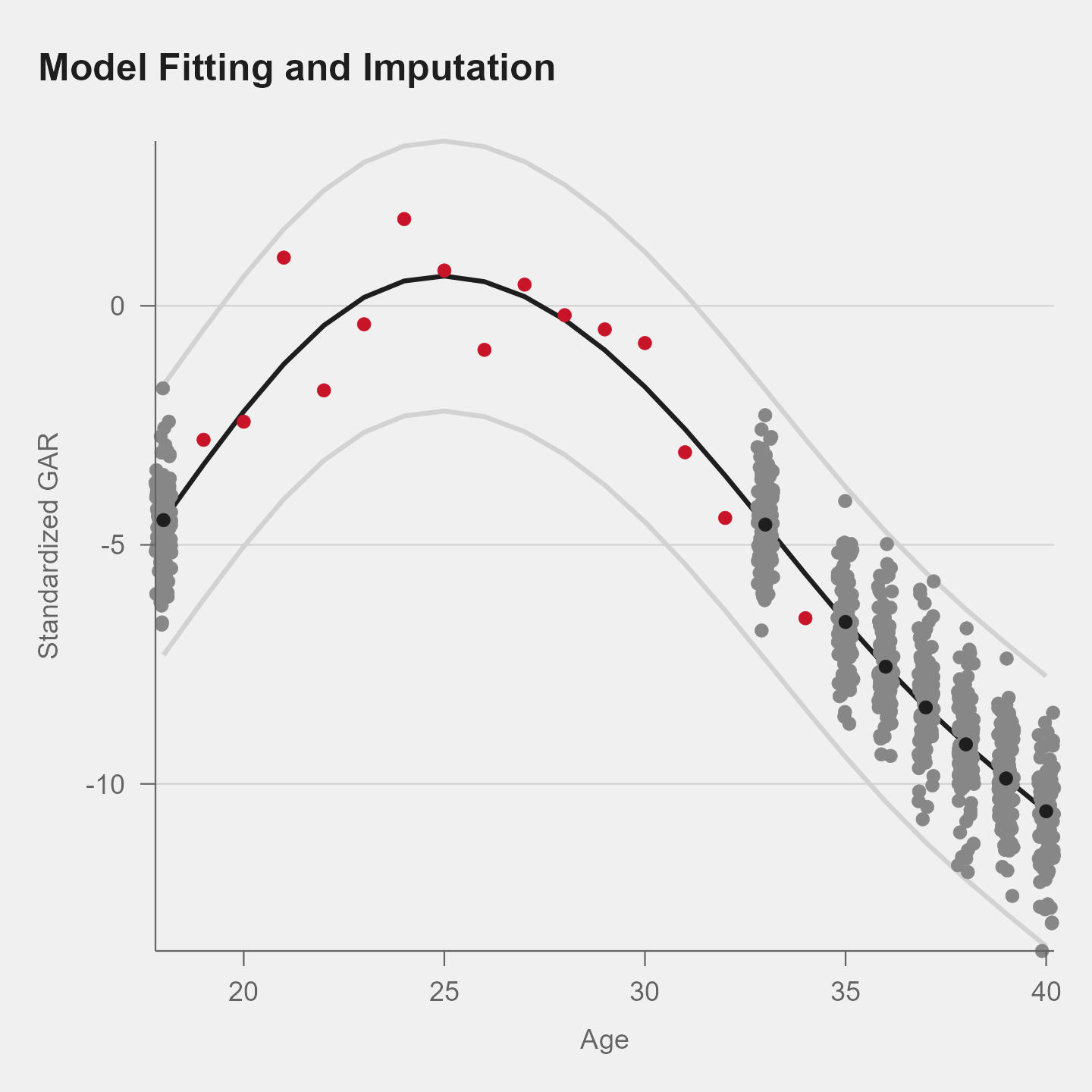}
    \includegraphics[width=.45\textwidth]{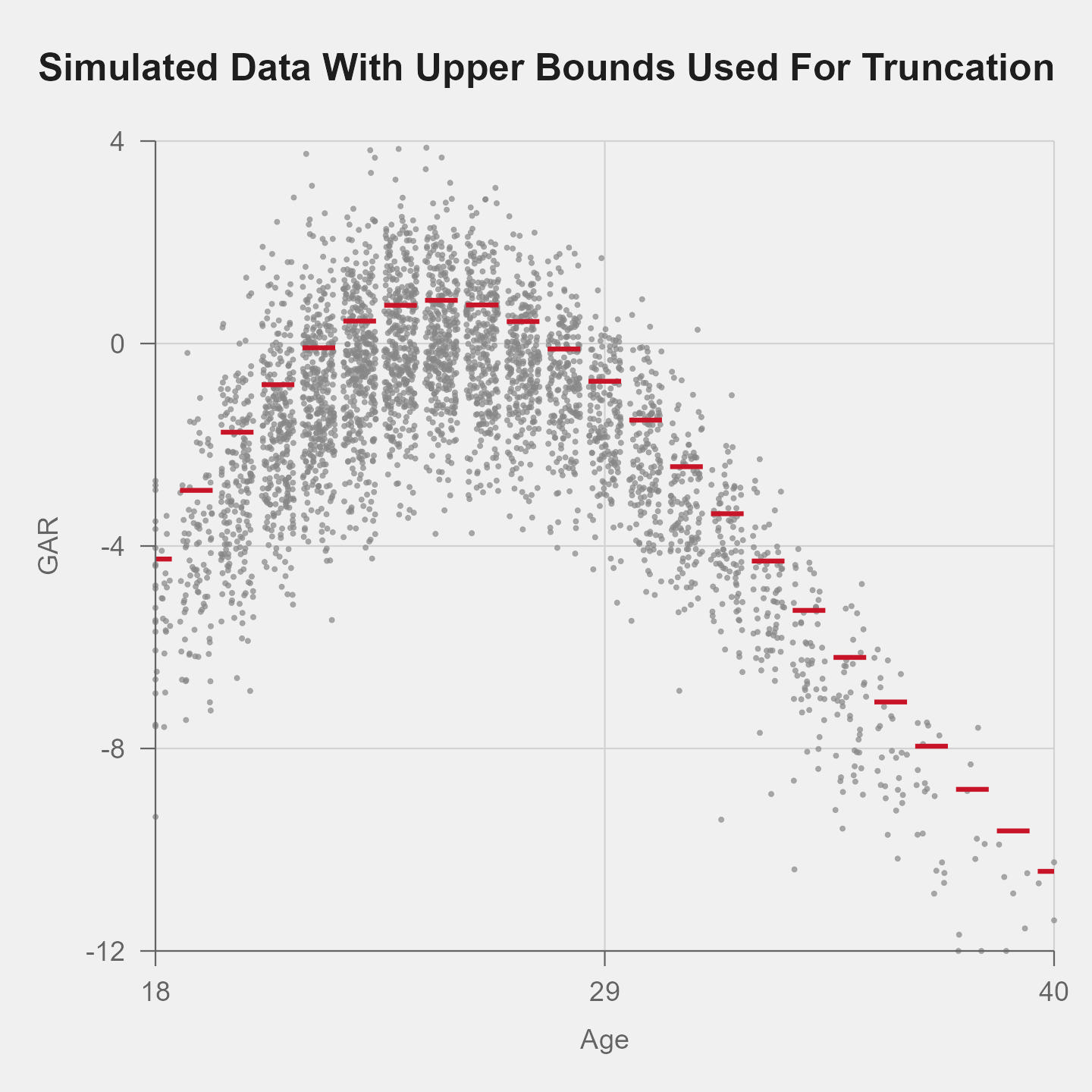}
    \caption{Modeling fitting and imputation without truncation for one example player in one simulation (left), and all simulated players (gray dots) and upper bounds (red line) used for truncation (right). In the left figure, red dots are observed data for one example player from one simulation, the black line denotes fitted values using the cubic spline model \textit{spline:trunc:fixed} for this player. The black dots, which lie on the line, are the means of the normal distribution that is used for imputing data for the ages for which data is missing, and the gray lines are 3 standard deviations away from the mean for those normal distributions.}
    \label{fig:modelfittingandimputation}
\end{figure}

Figure \ref{fig:modelfittingandimputation} depicts Steps 1 and 3, model fitting and imputation without truncation for one example player in one simulation (left) and the upper bounds used for truncation at each age for one of the simulations that did use truncation (right).  Our reasoning for  steps five and six is that there is possibly some initial effect in the model $\eta_{it}=\mu_{0t} + \gamma_{0i}$ which comes from fitting to \textit{only} fully observed data.   By adding the second iteration we hope to wash out some of that initial impact.  Our use of the boundary for imputation is based upon the idea that there are players who may be good enough to have their performance observed, that is play in whatever top league, but do have that opportunity, while at the same time there are players whose performance puts them well below the performance of other in the same league.  Additionally, the estimation of $\eta_{it}$ for generation of the imputed values was done two ways: using the quantile approach described above as well as the spline approach.

\subsection{Player Effects}
In addition to the basic model methodology and the data that we fit to these models we  considered some additional fixed and random effects for players.  The most common approach we used was a fixed constant player effect denoted as \textit{fixed}.  Our notation for a this fixed effect is $\gamma_{0i}$.  We considered two random effects models one with quadratic random effects and the other with spline random effects.  The former we denote by \textit{random-quad} and it has the following functional form: $g_{0i} + g_{1i} t + g_{2i} t^2$ \textcolor{black}{where each $g_{*i}$ is a random effect term for player $i$}.  For the spline random effects components, our shorthand is \textit{random-spline} and our function form is $g_{0i} + \xi_i(t)$.  
If the model did not have a fixed or random effect for player we use \textit{none} in our shorthand.  

In the subsections above, we have described three facets of our estimation approaches for $g(t)$.  We combine these three facets to consider a range of estimation methods though we do not consider all the possible combinations of these approaches.  We chose a specific subset of these combinations to make comparisons between our proposed methodologies and existing methodologies.  Table \ref{tab:estim} shows the full list of methods that we considered for evaluation along with shorthand and descriptions of the methodology.  For example, \textit{spline:trunc:fixed} uses 
natural splines to estimate $g(t)$ fit to both observed and truncated imputed data via a model that includes a fixed effect for each player.  Likewise \textit{quant:obs:none} use the quantile methodology described above to estimate the mean aging curve using only the observed $Y_{it}$'s.  Somewhat unique is \textit{quant:trunc:fixed} which generates the mean of the truncated imputed values via the quantile approach then fits a natural spline with fixed player effects to those truncated values.

\section{Simulation Study}

\subsection{Simulation Design}

We created a series of simulations to evaluate how well the various approaches described in the previous section estimate $g(t)$.  To generate data we followed a similar approach to that of \cite{tutoro2019} and created an underlying smooth curve, and generated values for player performance based upon that curve.  
We then modelled the dropout of players through a missingness process that generated values for $\psi_{it}$.
These simulations focused on the distribution of players and the methodology for their missingness.  

\subsection{Simulating individual player observations} Our simulated performance values for the $i^{th}$ player in year $t$ are generated in the following manner: 

\begin{equation} 
Y_{it} = \omega + \gamma_{0i} +a (t-t_{max})^2 + (b+b_i) (t-t_{max})^2 I_{\{ t> t_{max}\}} + c(t-t_{max})^3 I_{\{ t> t_{max}\}}+ \epsilon_{it} \label{eq:generatingcurve}
\end{equation}
%a=-1/9,b=-0.006,c=0.0045,agemax=25

\noindent where $i=1,\ldots, N$ and $t=t_1,\ldots,t_K$. The first three terms form a piecewise quadratic curves that serves as the underlying generating curve. The terms $\gamma_{0i} \sim N(0,\sigma^2_{\gamma})$ \textcolor{black}{ and $b_i \sim N(0,\sigma_b^2)$}. Thus this model is a piecewise cubic function with constant and quadratic player random effects. \textcolor{black}{ The constant player random effects are given by $\gamma_{0i}$ and the quadratic effects are given by $b_i$. The term $t_{max}$ represents the age at which maximal average performance is achieved.  The model is then quadratic up to ages of $t_{max}$ and cubic thereafter.  }

We can rewrite Equation \ref{eq:generatingcurve} to highlight the model components as:
\begin{eqnarray}
Y_{it}&=& g(t) + f(i,t) + \epsilon_{it},\\
g(t) &=& \omega  +a (t-t_{max})^2 + b (t-t_{max})^2 I_{\{ t> t_{max}\}} + c(t-t_{max})^3 I_{\{ t> t_{max}\}},\\
f(i,t)&=& \gamma_{0i} + b_i (t-t_{max})^2 I_{\{ t> t_{max}\}}. 
\end{eqnarray}
%to illustrate that our $g(t)$ is a piecewise cubic model with player effects that are quadratic.
%** I suggest we tie this back to the notation in \eqref{eq:model}, where the piecewise quadratic is $g(t)$, the intercept and player quadratic are $f(i,t)$ and the noise is $\epsilon_{it}$** is the simulated player differences from the mean curve which involves a player-specific intercept that simulates variations in overall player ability, and a player-specific quadratic term that simulates variations in how players age. 
The term $\epsilon_{it} \sim N(0, \sigma^2_{\epsilon})$ is random noise  that simulates year-to-year randomness in player performance and variation unaccounted for elsewhere in the model.   
For our simulations we fixed $t_{max}=25$, $a = -1/9$,  $b =-6/1000$, $c = 45/10000$ $t_{1}=18$, $t_{K} =40$, \textcolor{black}{$\sigma_b=0.02$} and $\sigma^2_{\epsilon}=1$.  Our choice of these values for $a$, $b$, and $c$ was based upon trying to match the \textcolor{black}{general} pattern \textcolor{black}{ of the percentage of players who were in the National Hockey League in a given year as well as those who transitioned from being in the league to being out of the league and vice versa in a given year.} 
%\textcolor{black}{\sout{of drop-in/drop-out for a subset of National Hockey League Forwards.}}  
We used varying values of $N$, $\omega$, and $\sigma_{\gamma}$ in our simulations.  In particular we simulated a full factorial from $N=300, 600, 1000$, $\omega = 0, 1$ and $\sigma_{\gamma} = 0.4, 0.8, 1.5$.  The values of $N$ that we chose were to illustrate the impact of sample size on estimation performance while also have a reasonable range of values.  For $\omega$ which represents the maximum value that our simulated $g(t)$ would take, we want to have both a zero and a non-zero option.  Finally, we based our choices for $\sigma_{\gamma}$ on the standard deviation of the estimated player effects from fitting a natural spline with player effects model to standardized points per game from National Hockey League forwards.  That value was approximately $0.8$ and we subsequently chose values for $\sigma_{\gamma}$ that were half and twice as large.  For each combination of values for $N$, $\omega$ and $\sigma_{\gamma}$, we simulated $200$ data sets and calculated the estimated $g(t)$ for each method on each of these data sets.

\begin{figure}
    \centering
    \includegraphics[width=.9\textwidth]{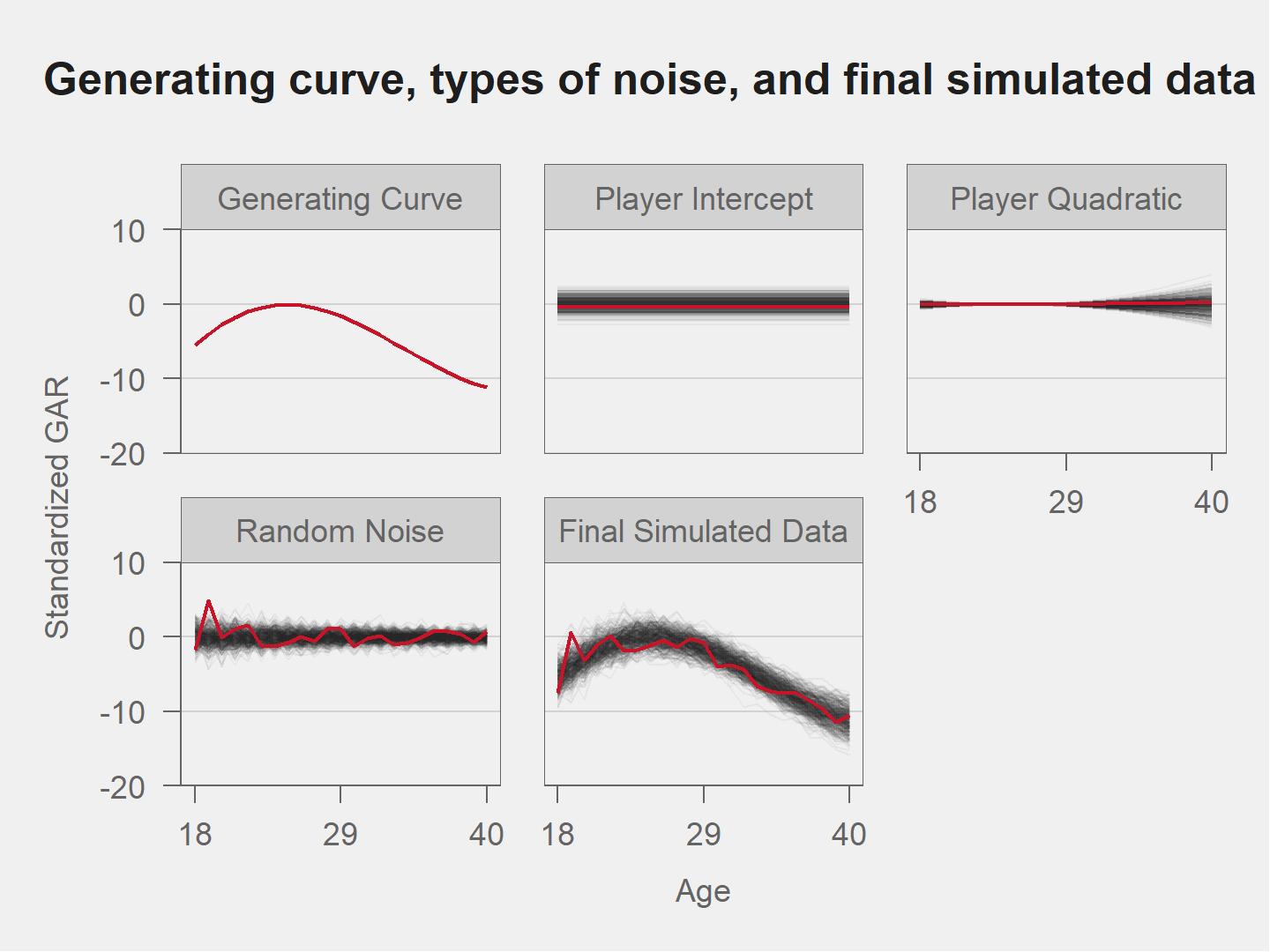}
    \caption{The underlying generating curve, different types of noise, and the final simulated data used in one simulation. The gray curves represent simulated players.  The curve for one simulated player is highlighted in red.}
    \label{fig:simulatingplayercurves}
\end{figure}

For example, the components of $g(t)$ from \eqref{eq:generatingcurve} are depicted in Figure \ref{fig:simulatingplayercurves}. Shown in that figure are the underlying generating curve, the player intercept terms for overall player ability, the player quadratic curves for player aging, the random noise terms, and the final simulated player curves. These were taken from one simulation with the parameters $N=600$, $\omega=0$, and $\sigma_{\gamma}=0.8$.  All simulated players are shown in gray, and the curves generated for one particular simulated player are shown in red. This highlighted player is slightly below average,  intercept is slightly below zero, and ages slightly more slowly than average.  The player quadratic is slightly concave upward, which when added to the generating curve, gives a curve that indicates that this player's decline will not be as steep as an average player. 

\subsection{Simulating missingness of observations} 
For generating the $\psi_{it}$ values for each player and age, we chose a methodology that tried to %\textcolor{black}{\sout{mirror} approximate} 
the \textcolor{black}{aging trend and the percent of players observed at a given year for}  forwards in the National Hockey League.  Recall that $\psi_{it} = 1$ if observation $Y_{it}$ was observed.  Using the methodology from the previous section, we simulated $Y_{it}$ for each player $i$ and each age $t$ between $t_1=18$ and $t_K=40$.  Using those values, we obtained $\psi_{it}$ for each player via the following steps:
\begin{enumerate}
    \item For each age $t$, round $N \pi_t$ to a whole number and call that $n^{+}_t$.
    \item For each player $i$, calculate $\rho_{it} = \exp \{\sum_{k=1}^{t} Y_{ik}\}.$
    \item Sample without replacement $n^{+}_t$ players from the $N$ players with each having
    probability of selection $p_{it} = \frac{\rho_{it}}{\sum_{j}\rho_{jt}}.$
    \item Make $\psi_{it} =1$ for the $n^{+}_t$ players selected in the previous step and $\psi_{it}=0$ for the remaining players.
\end{enumerate}
Simulations that used the cumulative performance missingness generation were paired with a full factorial of values for $\omega = 0, 1$, $N=600, 800, 1000$,  and $\sigma_{\gamma} = 0.4,0.8, 1.5$.  For simplicity we assumed that all simulated value of $Y_{it}$ were observable.
%\sout{, i.e.~ $\phi_{it}=1$ for all $i$ and $t$}.
To evaluate the methods proposed in the previous section, we generated simulated data following the data generation and missingness approaches described above in this section.

%Below we discuss simulations for each combination of parameters we generated $200$ sets of player data and fit each of the methods above for estimating the average player aging curve.  For evaluation of the performance methods, we introduce $$\mu_t^{\dagger} = \frac{1}{n^{\dagger}_t} \sum_{i \in \Omega  \& \psi_{it}<K} Y_{it}$$  
%where $\Omega = \{ i \mid \sum_i \psi_{it} >0 \}$ is the collection of players who had at least one observed performance value and
%$n^{\dagger}_t = \sum_{i \in \Omega} \psi_{it}$ is the number of players at age $t$ whose performance was observed.  Our focus is on $\mu_t^{\dagger}$ since it is possible given our missingness generation process that not all players will have at least one observed value.  Note that, in general, since our missingness generation makes it likely that players with lower values for $Y_{it}$ are not observed --- just like actual sports --- , then $\mu_t \leq \mu_t^{\dagger}$.  Because of this potential bias, we will judge the performance of our estimation methods against $\mu_t^{\dagger}$.

\begin{figure}
    \centering
    \includegraphics[width=.75\textwidth]{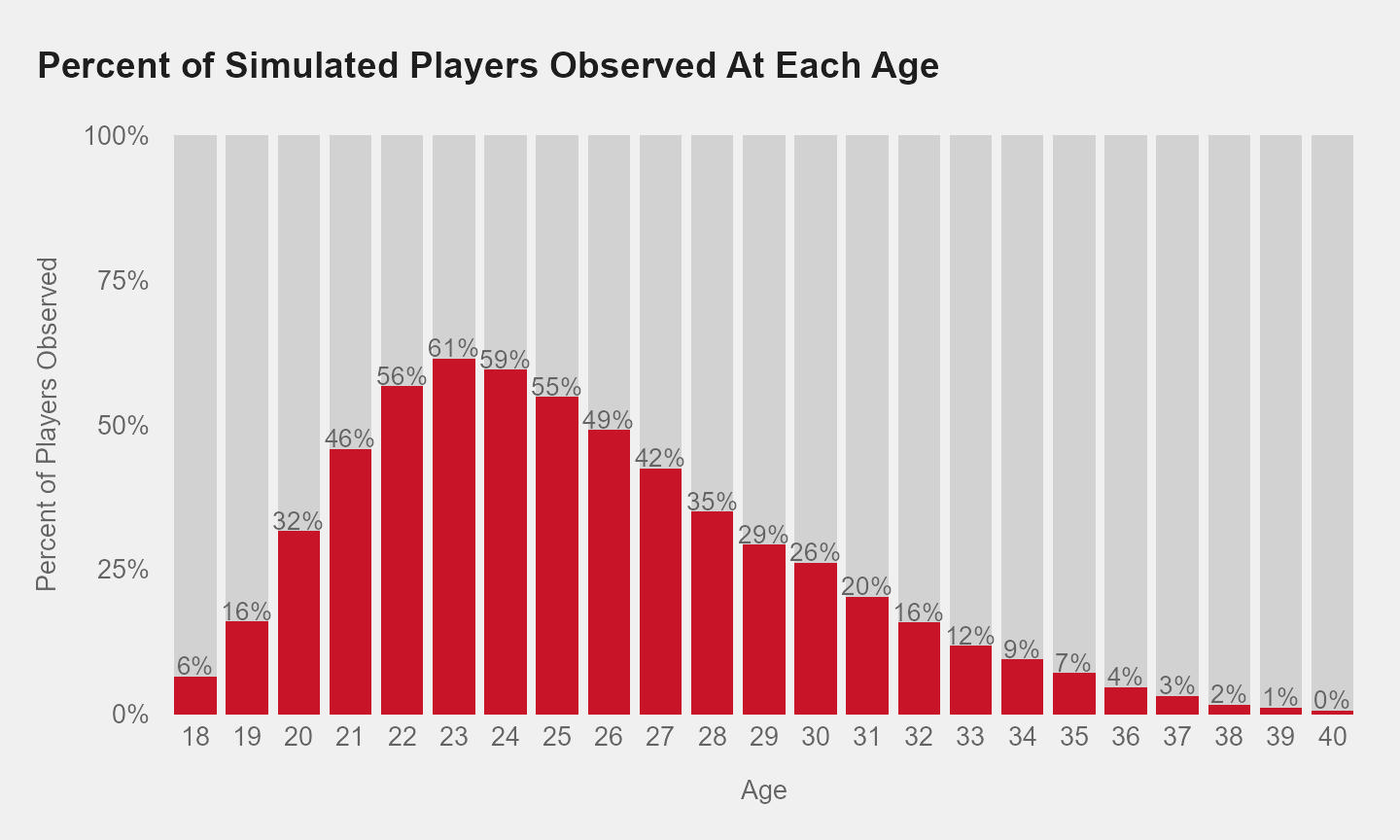}
    \caption{Percent of players that are observed at each age for the simulated data. The distributions follows similar shape as \ref{fig:percentobservednhl} and peak around ages 23-24.}
    \label{fig:percentobservedsim}
\end{figure}

Figure \ref{fig:percentobservedsim} shows the percentage of players that are observed by age for NHL forwards (left) and for simulated data (right) with $N=600$, $\omega = 0$ and $\sigma_{\gamma}=0.8$.

\subsection{Simulation Results}
\label{sec:evaluation}
Next we consider how each of our methods performed in their estimation of the mean aging curve on their average error and on their overall shape.  

\subsubsection{Root Mean Squared Error}

\begin{figure}
    \centering
    \includegraphics[width=14cm,
  keepaspectratio]{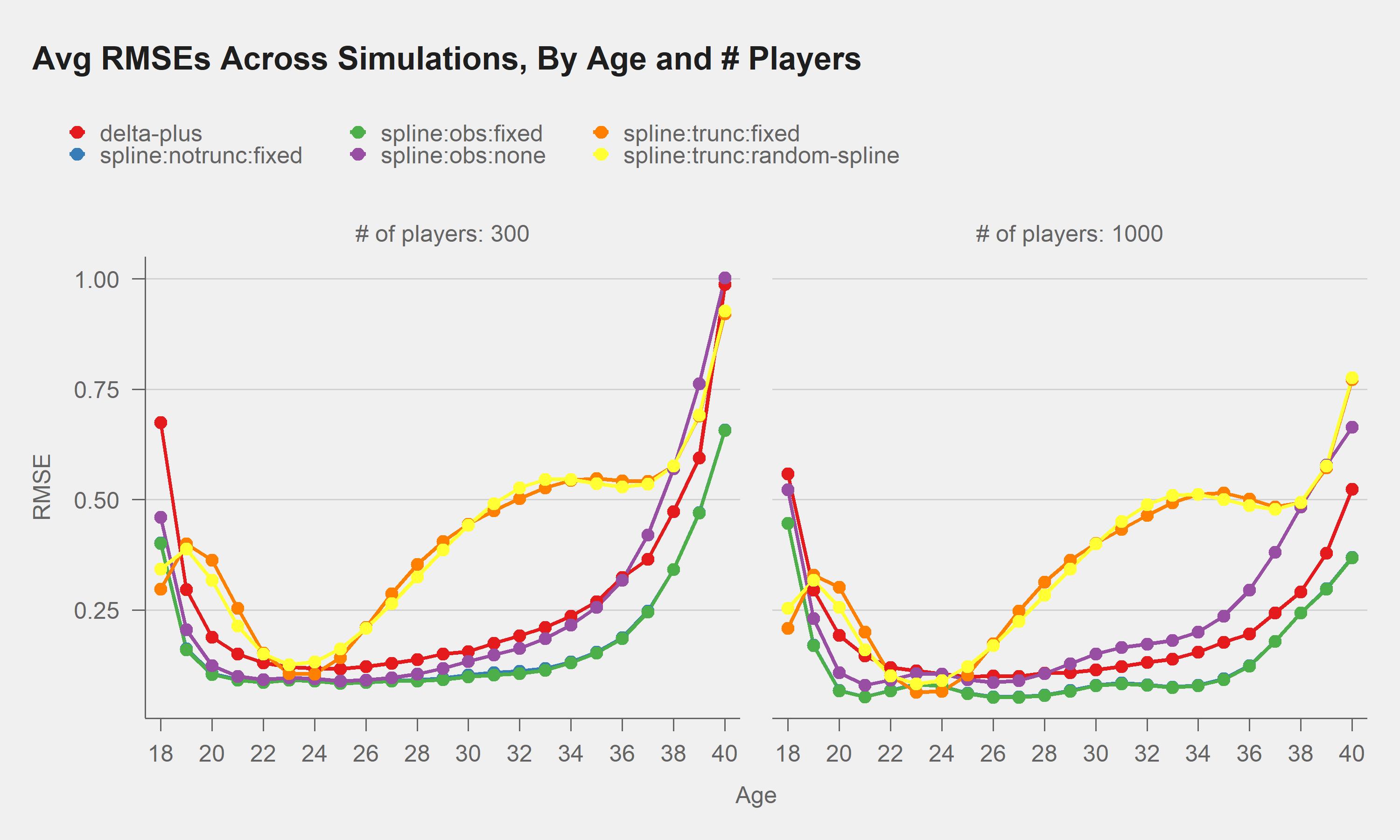}
    \caption{Root Mean Squared Error by Age and Number of Players.}
    \label{fig:rmse_age_numbplayers}
\end{figure}

Estimated curves are first compared to $g(t)$ on their average root mean squared error (RMSE) at each age.  Figure \ref{fig:rmse_age_numbplayers} shows RMSE by age, averaged across simulations, and split for each of simulations with 300 versus 1000 players. Six curves are shown; ones not presented in Figure \ref{fig:rmse_age_numbplayers} showed RMSE's that were larger and required a re-scaling of the y-axis that rendered comparisons of the remaining methods difficult.

RMSE at each age is lowest at around age 24, which corresponds to when the majority of NHL players in Figure \ref{fig:rmse_age_numbplayers} are observed. At entry (age 18) and at the end of careers (age 30 onwards), RMSE's tend to be higher. 

Overall, two methods -- \textit{spline:obs:fixed} and \textit{spline:notrunc:fixed} --- boast the lowest average RMSEs at each age. This suggests that for estimating age curves, either imputation or player specific intercept models are preferred.  These two methods yield nearly identical average RMSEs at each age, which is why, the \textit{spline:notrunc:fixed} (blue) method is not visible in Figure \ref{fig:rmse_age_numbplayers}.

All models better estimate age curves with the larger sample size (the right facet of Figure \ref{fig:rmse_age_numbplayers}). The impact of sample size is largest for the   \textit{delta-plus}  method.  For example, at age 40, \textit{delta-plus}  shows the 3rd lowest RMSE with 1,000 players, but the 5th lowest RMSE with 300 players. Additionally, \textit{delta-plus} is worse with the lowest age group -- for both 300 and 1000 simulated players, it shows the largest RMSE in Figure \ref{fig:rmse_age_numbplayers} among players at age 18. 
In general, simulations with higher standard deviations (1.5, versus 0.8 and 0.4) averaged higher RMSEs, although the impact of increasing standard deviation appeared uniform across age curve estimating method.

\subsubsection{Shape Based Distance}

To supplement simulation results based on RMSE, we use shape based distance (SBD) \cite{paparrizos2015k}, a metric designed to approximate the similarity of time series curves. SBD normalizes the cross-correlation between time series curves, which in our example is the estimated age curve and the truth. SBD scores ranges from 0 to 1, where lower scores reflect more similar curves. 

\begin{figure}
    \centering
    \includegraphics[width=12cm,
  keepaspectratio]{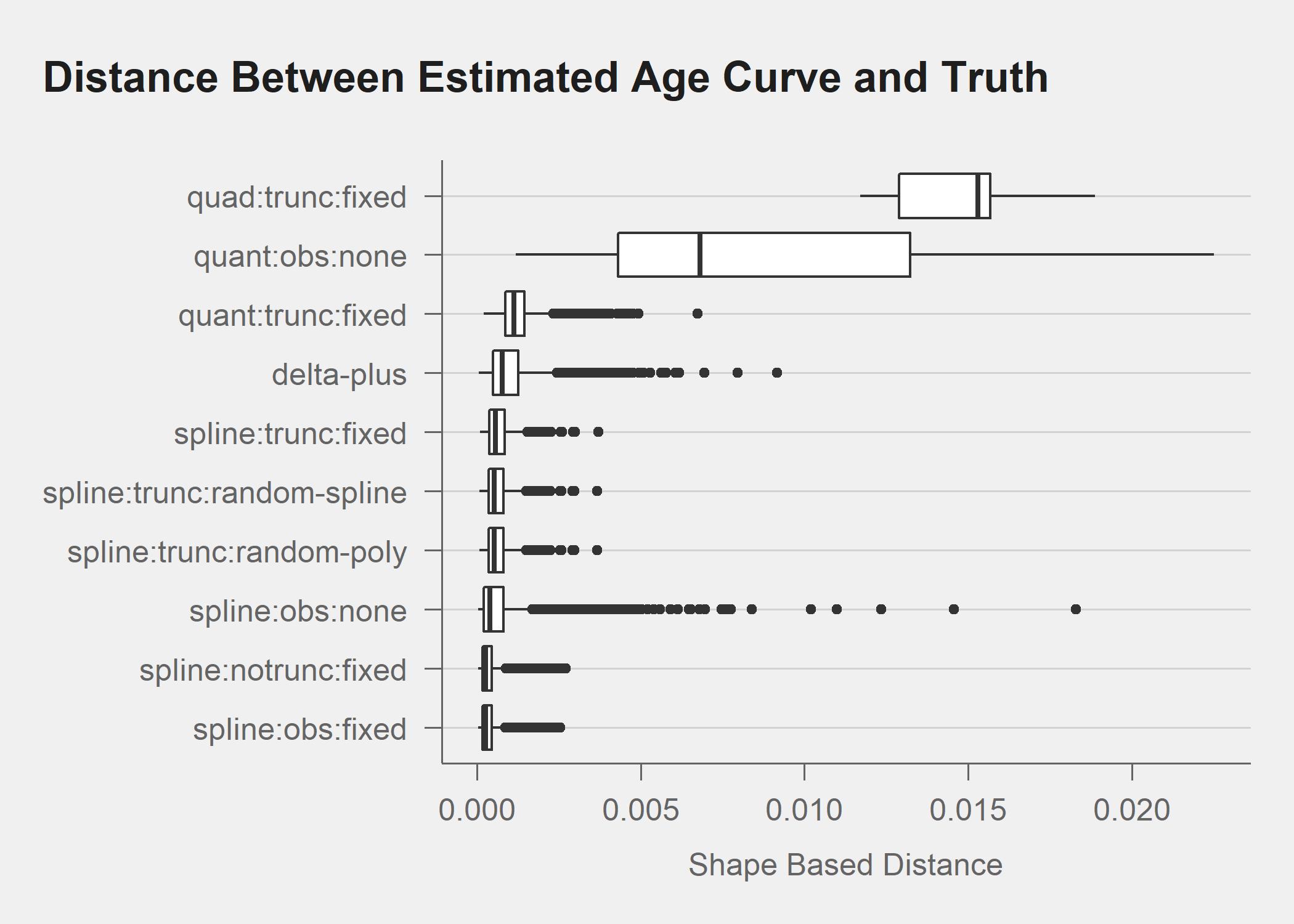}
    \caption{Shape Based Distance between true and estimated age curves, across simulations}
    \label{fig:sbd}
\end{figure}

Figure \ref{fig:sbd} shows a boxplot with shape based distances at each simulation for each method.  
As in Figure \ref{fig:rmse_age_numbplayers},
\textit{spline:obs:fixed} and \textit{spline:notrunc:fixed}
 boast curve shapes that are closest to the true age curve in Figure \ref{fig:sbd}. The basic spline model without a player intercept (\textit{spline:obs:fixed}) has the third lowest median SBD; however, this method also boasts several outlying SBD observations. Similarly, the \textit{delta-plus} shows SBD outliers. The method that assumes a quadratic response surface has the highest median SBD, and is in the top row of Figure \ref{fig:sbd}, suggesting that quadratic-based approaches may not identify the shape of age curves. 

\section{Application to National Hockey League Data}

To illustrate the impact of the methods proposed in this paper, we apply some of the approaches in this paper to data on \textcolor{black}{2276} NHL forwards who were born on or after January 1, 1970.  The data were obtained from \url{www.eliteprospects.com} and only players who played at least one season in the National Hockey League were included.  
For our measure of player performance, $Y_{it}$, we chose a standardized points (goals plus assists) per game where the z-score was calculated relative to the mean and standard deviation for the NHL in that particular season.  Our selection of this metric was based upon its availability for all players across the range of seasons (1988-89 season to 2019-20 season).

In Figure \ref{fig:estimatedagecurves} we show estimated age curves for NHL player points per game using different methods. The differences in our two best methods, \textit{spline:notrunc:fixed} (red) and \textit{spline:obs:fixed} (lightred), neither of which use truncation, are not perceptible as their lines overlap. The methods with truncation, \textit{spline:trunc:fixed} (blue), \textit{spline:trunc:random-quad} (gray) and 
\linebreak[4] 
\textit{spline:trunc:random-spline} (black) are very similar and have overlapping curves as well. Those curves are lower than the curves from our best methods, which is expected since imputed values were taken from a truncated normal.   The overall pattern for all of these curves does not seem quadratric.  
%The delta-plus (red) and quadratic model quad:trunc:fixed (blue) are noticeably different, especially for higher ages. The jaggedness of the delta-plus age curve at higher ages is due to small sample size issues for higher ages.  The spline model using only observed data and no player effect (spline:obs:none, lightred) is shown as well for comparison. 
\begin{figure}
    \centering
    \includegraphics[width=.45\textwidth]{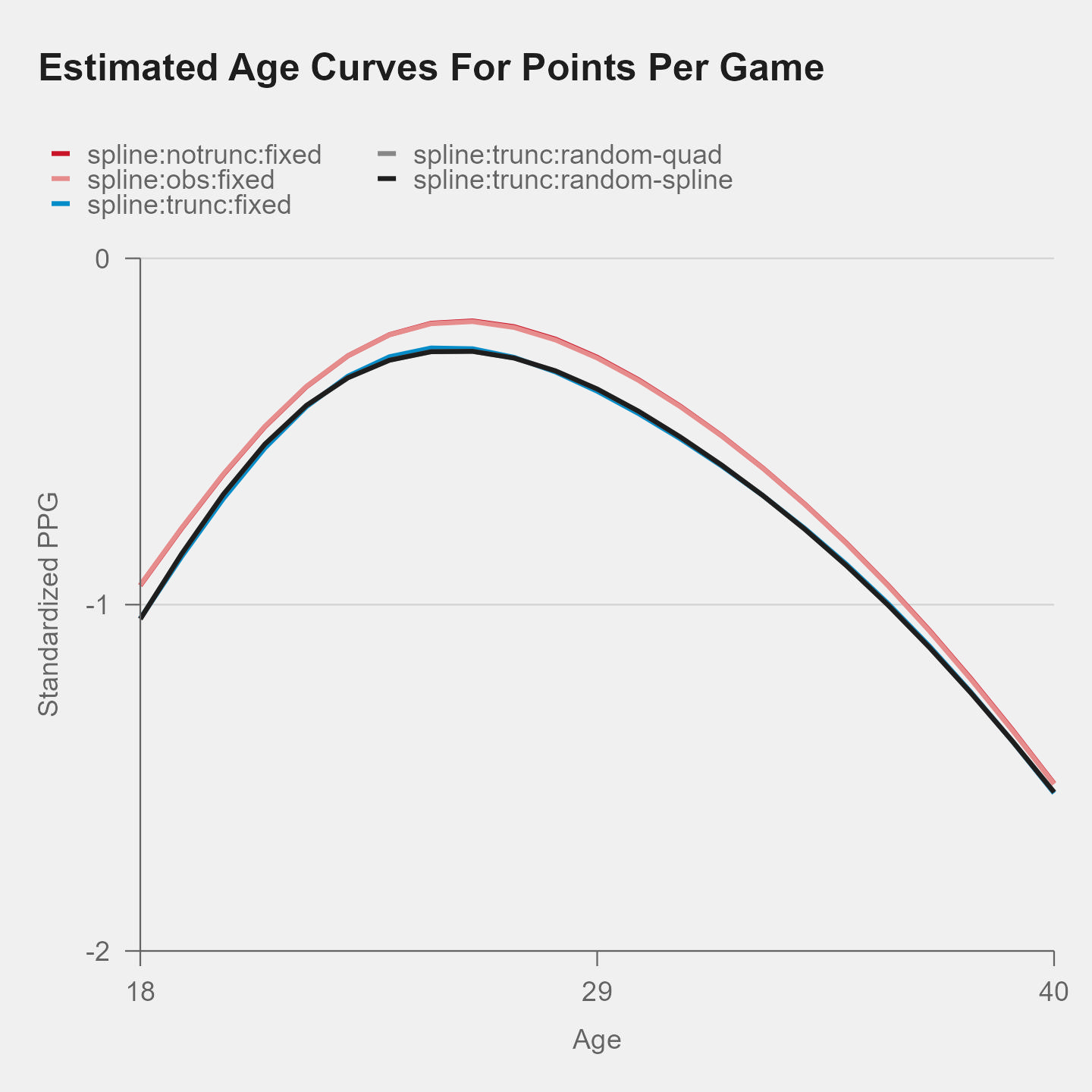}
     \includegraphics[width=.45\textwidth]{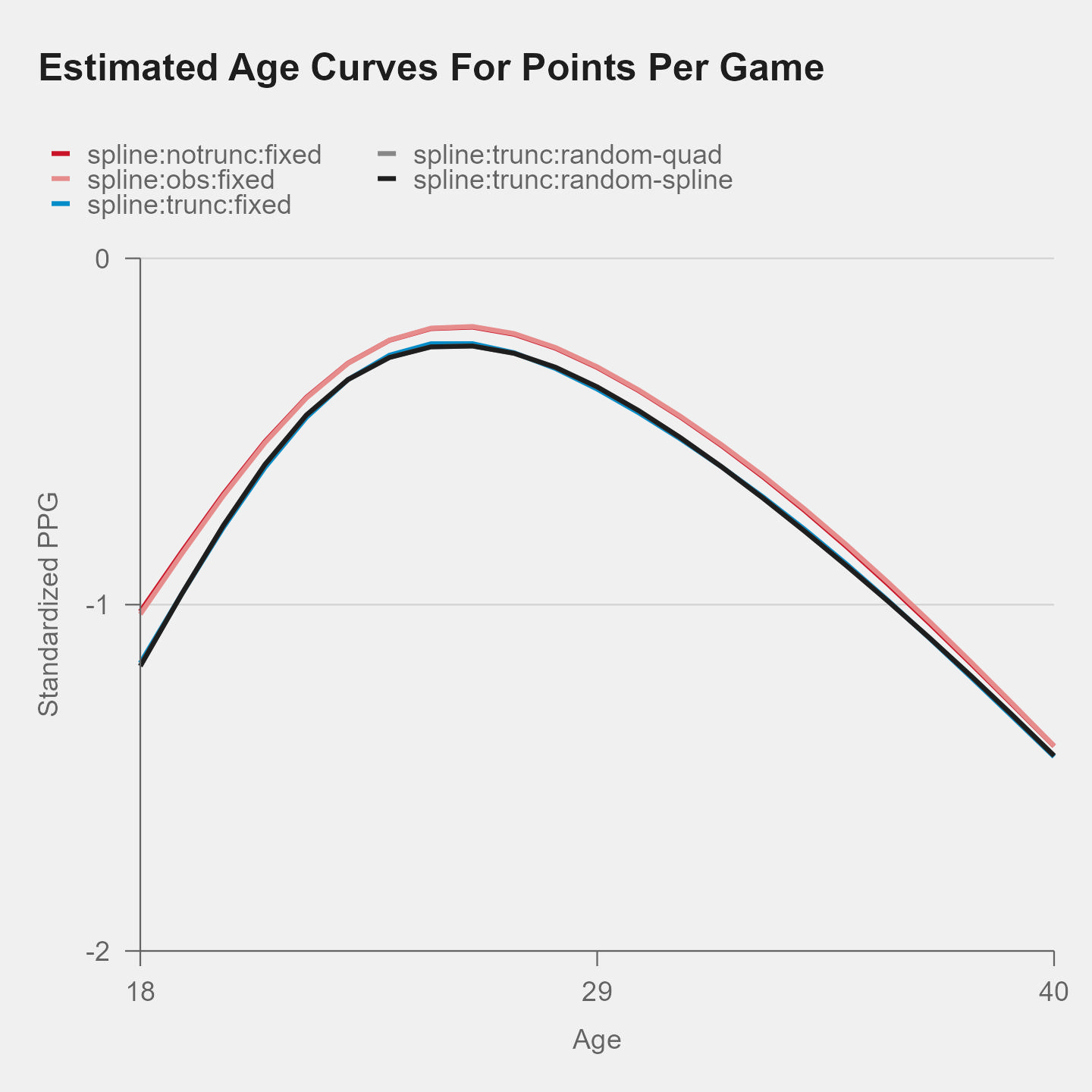}
    \caption{Estimated age curves for NHL points per game  using different methods.} 
    \label{fig:estimatedagecurves}
\end{figure}

\section{Discussion}

Understanding how player performance changes as players age is important in sports, particularly for team management who need to sign players to contracts.  
In this paper we have proposed and evaluated several new methodologies for 
estimation of mean player aging curves. In particular, we have presented formal methods for adding imputed data to augment the missingness that regularly appears in data from some professional sports leagues.   The models that performed best in our simulation study had either a flexible form or incorporated player effects (through imputation, or directly).  This paper has also presented a framework for incorporating imputed data into estimation of player aging effects.  With player effects, the age curve term contains information about the relative changes in performance from age to age and overall performance is part of the player term.  When estimation is done with only observed data, the relative changes in the mean aging curve are only utilized where a particular player is present.  Without player effects as a model component, the individual player and the overall are entangled.  Specifically, we find that a spline methodology with fixed player effects allow perform better than other methods and these methods have the flexibility to appropriately and efficiently estimate player aging curves.

There are some additional considerations that might improve the performance of the methods we have considered.  One possibility would be to consider a fully Bayesian approach that treats all of the unknown aspects of the model both the $g(t) + f(i,t)$ and unobserved $Y_{it}$'s as random variables.  This approach could consider a complete posterior inference given the uncertainty in estimation.  Along similar lines, an approach that does multiple imputation for each unobserved $Y_{it}$ could improve performance.   Another assumption that has been made in the literature is to treat ages as whole numbers.  It certainly seems possible for a regression based approaches for estimation of $g(t)$ to deal in fractional ages $t$ for a given season.

Our simulation study was focused on \textcolor{black}{generating} mean aging curves \textcolor{black}{and player-age missingness} that yielded similar \textcolor{black}{
%\sout{player drop-in/drop-out} 
aging and missingness} patterns to those observed in a particular professional league, the National Hockey League. \textcolor{black}{Compare Figure \ref{fig:percentobservednhl} and Figure  \ref{fig:percentobservedsim}}.  It would certainly be reasonable to consider other functional forms for the underlying $g(t)$ and $f(i,t)$, though we believe that the results from such a study would be in line with those found in this paper.  Another avenue of possible future work would be to consider additional mechanisms for generation of the $\psi_{it}$ values. 

Overall the novel methods proposed and evaluated in this paper via simulation study have improved our understanding of how to estimate player aging curves.  It is clear from the results in this paper that the best methods for estimation of player aging are those that have model flexibility and that include player effects.  The use of imputation also has potential to impact this methodology and, thus,  being aware of what we don't observe can make our estimation stronger.

%- an approach that uses $s_t$ as estimated standard deviation per age.

%That's prob not the greatest way of saying it. Maybe an example like this would help illustrate what's going on

\begin{comment}
\begin{enumerate}
    \item Model Without Player effect: Given age is 36, E(PPG) is xxxx (bigger than it should be because only good players are left at age 36). The E(PPG) is an average over observed players.

    \item Model With  Player effect: 
    \begin{enumerate}
        \item Given age is 36 AND player is Wayne Gretzky, E(PPG) is xxxx.  
        \item Given age is 36 AND player is (an average observed/unobserved player whose random intercept is 0), E(PPG) is (something lower). 
    \end{enumerate}
\end{enumerate}
In (1), E(PPG) is an average over observed players, and in (2), the age curve term in E(PPG) is an average over observed/unobserved players. 2(b) is basically the age curve term in \verb'loess_nomiss', \verb'spline_nomiss' models, and that is lower for older ages than the curve in (1).
\end{comment}

\begin{acknowledgements}
We would like to thank CJ Turturo for making his code available.  
\end{acknowledgements}

 \section*{Conflict of interest}

 The authors declare that they have no conflict of interest.

\newpage

\section{Appendix}

Below is a table highlighting average RMSE for different simulation settings, by age. 

\begin{figure}[h]
    \centering
    \includegraphics[width=1.0\textwidth]{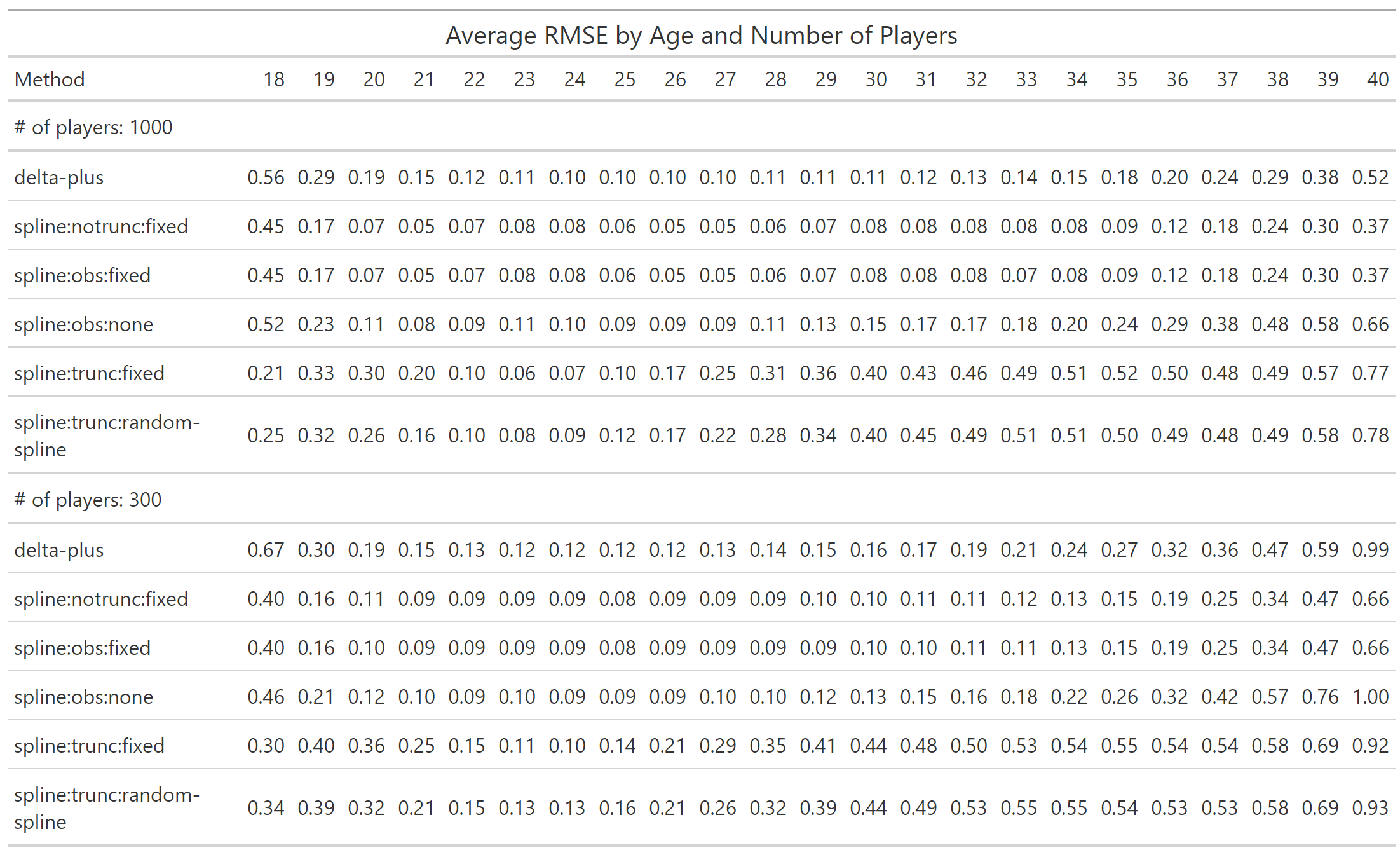}
    \caption{Average Root Mean Squared Error by Age (column), Method (row), and number of players (sets of rows) across simulations.} 
    \label{fig:estimatedagecurves}
\end{figure}

\begin{comment} 
 \begin{table}[]
    \centering
    \caption{Summary of Estimation Methods}
    \begin{tabular}{l|l|c|c}
    \hline
         Method Name & Model formulation &Estimation of $g(t)$ & Data  \\
         \hline
 Delta-plus (delta-plus) & Non-parametric & Piecewise & No\\
 Spline\underline{ }noplayer\underline{ }nomiss (spline:obs:none)&
    $g(t)$ & Natural Splines & No\\
  Spline\underline{ }nomiss
  (spline:obs:fixed)& $g(t) + \gamma_{0i}$& Natural Splines & No\\
  Miss1 (spline:trunc:fixed)& $g(t)+\gamma_{0i}$& Natural Splines & Yes (truncated)\\
   Miss\underline{ }notrunc  (spline:notrunc:fixed)&$g(t)+\gamma_{0i}$& Natural Splines & Yes (not truncated)\\
  Miss2 (quant:trunc:fixed)& $g(t)+\gamma_{0i}$ & Quantile Approx. & Yes (truncated) \\
  Miss\underline{ }quantile (quant:obs:none)& $g(t)+\gamma_{0i}$ &Quantile Approx. & No\\
  Miss\underline{ }quadratic
  (quad:trunc:fixed)& $g_0+g_1 t + g_2t^2 + \gamma_{0i}$& Quad. Linear Model& Yes (truncated)\\
  Quad\underline{ }random (spline:trunc:random-quad)&$g(t)+\gamma_{0i} + \gamma_{1i} t + \gamma_{2i} t^2$ & Natural Splines & Yes (truncated)\\
  Spline\underline{ }ns2 (spline:trunc:random-spline)& $g(t) +\gamma_{0i} + \xi_i(t)$ & Natural Splines & Yes (truncated)\\
    \hline
    \end{tabular}
\end{table}
\end{comment}

\newpage
 
\bibliographystyle{spmpsci}   
\bibliography{Ref} 

\newpage
\end{document}